\documentclass[twocolumn,superscriptaddress,amsmath,amssymb,aps,pra,floatfix]{revtex4-1}

\usepackage[utf8]{inputenc}
\usepackage{wrapfig}
\usepackage{float}
\usepackage{amssymb}
\usepackage{amsfonts}
\usepackage{amsmath}
\usepackage{cases}
\usepackage{stmaryrd}\usepackage{tipa}
\usepackage[dvips]{graphicx}
\usepackage{dcolumn}
\usepackage{bbold}
\usepackage{graphicx}
\usepackage{graphicx}
\usepackage{subfigure}
\usepackage{dcolumn}
\usepackage{bm}
\usepackage{color}
\usepackage{xcolor}
\usepackage{CJK}
\usepackage{subfigure}
\usepackage{physics}
\usepackage{array}
\usepackage{multirow}
\usepackage{nopageno}
\usepackage{mathtools}
\usepackage{enumerate}
\usepackage{comment}
\usepackage[colorlinks=true, linkcolor=red, citecolor=blue]{hyperref}
\usepackage{diagbox}
\usepackage{ulem}
\usepackage{makecell}
\setcellgapes{10pt}

\definecolor{gg}{RGB}{8, 135, 68}

\newcommand*{\colorboxed}{}
\def\colorboxed#1#{%
  \colorboxedAux{#1}%
}
\newcommand*{\colorboxedAux}[3]{%
  \begingroup
    \colorlet{cb@saved}{.}%
    \color#1{#2}%
    \boxed{%
      \color{cb@saved}%
      #3%
    }%
  \endgroup
}

\begin{document}

\title{Satellite-based entanglement distribution and quantum teleportation with continuous variables}

\author{Tasio Gonzalez-Raya}
\email{tasio.gonzalez@ehu.eus}
\affiliation{Department of Physical Chemistry, University of the Basque Country UPV/EHU, Apartado 644, 48080 Bilbao, Spain}
\affiliation{EHU Quantum Center, University of the Basque Country UPV/EHU}
\author{Stefano Pirandola}
\affiliation{Department of Computer Science, University of York, York YO10 5GH, United Kingdom}
\author{Mikel Sanz}
\affiliation{Department of Physical Chemistry, University of the Basque Country UPV/EHU, Apartado 644, 48080 Bilbao, Spain}
\affiliation{EHU Quantum Center, University of the Basque Country UPV/EHU}
\affiliation{IKERBASQUE, Basque Foundation for Science, Plaza Euskadi 5, 48009 Bilbao, Spain}
\affiliation{Basque Center for Applied Mathematics (BCAM), Alameda de Mazarredo 14, 48009 Bilbao, Spain}

\begin{abstract}
Advances in satellite quantum communications aim at reshaping the global telecommunication network by increasing the security of the transferred information. Here, we study the effects of atmospheric turbulence in continuous-variable entanglement distribution and quantum teleportation in the optical regime between a ground station and a satellite. More specifically, we study the degradation of entanglement due to various error sources in the distribution, namely, diffraction, atmospheric attenuation, turbulence, and detector inefficiency, in both downlink and uplink scenarios. As the fidelity of a quantum teleportation protocol using these distributed entangled resources is not sufficient, we include an intermediate station for either state generation, or beam refocusing, in order to reduce the effects of atmospheric turbulence and diffraction, respectively. The results show the feasibility of free-space entanglement distribution and quantum teleportation in downlink paths up to the LEO region, but also in uplink paths with the help of the intermediate station. Finally, we complete the study with microwave-optical comparison in bad weather situations, and with the study of horizontal paths in ground-to-ground and inter-satellite quantum communication.
\end{abstract}

\maketitle

\section{Introduction}

The advantages forecasted by quantum information theory spurred the development of quantum communications~\cite{Gisin2007,Yuan2010,Krenn2016,Pirandola2020}. A staple of quantum communication is quantum teleportation~\cite{Bennett1993,Braunstein1998,Pirandola2015}, a protocol that aims at transmitting the information contained in an unknown quantum state held by one party, to another party, by means of an entangled quantum resource that they both share. The latter occupies the area of entanglement distribution~\cite{Cirac1997,Chou2007,Mista2009}.

Recent experiments in quantum teleportation with optical fibres showcase a hard limit at 100 km~\cite{Takesue2015,Valivarthi2016,Sun2016,Huo2018}, mainly due to the inefficiency of single-photon detection. This limit was extended to 143 km when switching to free space~\cite{Ma2012}, connecting two ground stations; when linking a ground station with a satellite, quantum teleportation has been performed over 1000 km~\cite{Ren2017}. Similar distances were achieved for entanglement distribution in ground-to-ground~\cite{Yin2012} and ground-to-satellite~\cite{Yin2017} scenarios.

Setting aside the technological overhead, improvements will go through understanding the different loss mechanisms in free space, namely diffraction, atmospheric attenuation, and especially turbulence effects. The latter have been well studied for classical signals~\cite{Fante1975,Fante1980}, and some recent works have studied them in quantum atmospheric transmission channels~\cite{Vasylyev2016}, establishing that non-classicality of signals can be preserved~\cite{Vasylyev2012}. Insightful papers into the effects of free space propagation of quantum signals were published in Refs.~\cite{Liorni2019,Vasylyev2019}. More recently, others focused on the limits for key generation and entanglement distribution between ground stations~\cite{Pirandola2021_2} and between ground stations and satellites~\cite{Pirandola2021}. All these works have contributed to studying the limitations for the involvement of satellites in quantum communications~\cite{Pirandola2020,Sidhu2021}. Here, we contribute by studying how entanglement is degraded between ground stations and satellites for performing quantum teleportation. 

\begin{figure}[b]
{\includegraphics[width=0.45 \textwidth]{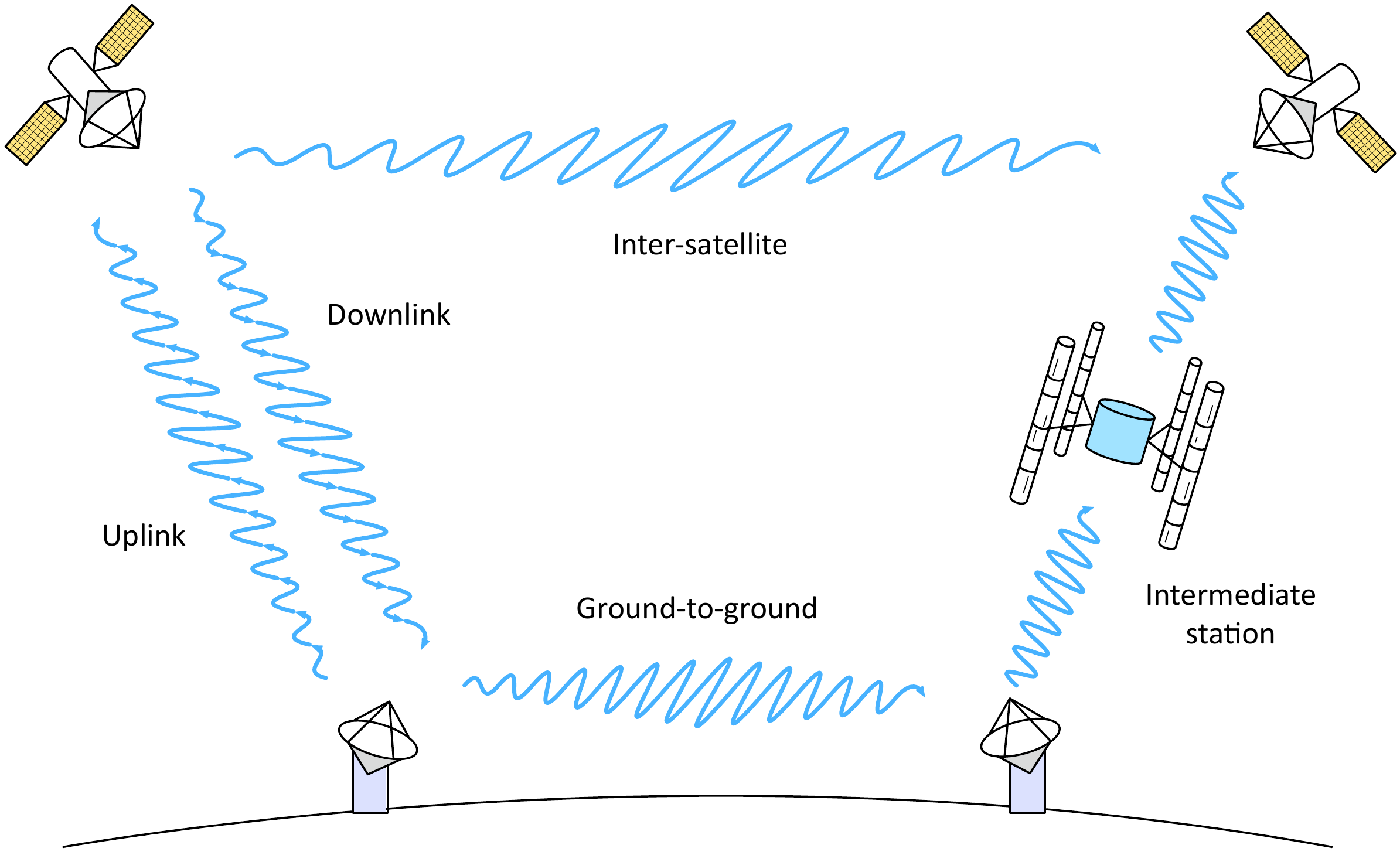}}
\caption{Sketch describing the different quantum communication scenarios that we have studied in this paper. We have investigated downlink and uplink channels, between a ground station and a satellite, both directly and with an intermediate station. We have also studied horizontal paths, between two ground stations and between two satellites.}
\label{fig1}
\end{figure}

More precisely, in this article, we investigate the effect of free-space turbulence on the propagation of quantum states in the optical regime, and how entanglement (quantified by the negativity of the covariance matrix) is degraded in this process. We assume that two parties attempt to share an entangled state, distributed through open air, to perform quantum teleportation. Then, we make use of the Braunstein-Kimble teleportation protocol~\cite{Braunstein1998} for continuous-variable (CV) systems. Particularly, we consider that we initially have two-mode squeezed vacuum (TMSV) states which are distributed through free space, and we use them to teleport a coherent state. We investigate different instances of quantum communication: ground station to satellite (uplink), satellite to ground station (downlink), and the placement of an intermediate station (intermediate), either to generate states, or to refocus the beam. The different quantum communication scenarios studied here are depicted in Fig.~\ref{fig1}.

We find that downlink communication from the LEO region leads to non-classical teleportation fidelities, while uplink paths of similar distances require an intermediate station to reduce the effect of atmospheric degradation of entanglement. We then study the limits for entanglement distribution and quantum teleportation with microwave signals, and compare them with optical signals, in a bad weather situation. We observe that the distances are highly reduced due to diffraction and thermal noise, as expected for microwaves. We conclude by investigating entanglement distribution and quantum teleportation in horizontal paths, i.e. between two ground stations and between two satellites.

\section{Theoretical framework}
We consider that we have a ground station at altitude $h_{0}$, and a satellite with an orbit radius $R_{0}$ and distance from the surface of the Earth $h\equiv R_{0}-R_{\text{E}}$, where $R_{\text{E}}$ is the radius of the Earth. Then the distance between the ground station, that sees the satellite at an angle $\theta$, and the satellite is
\begin{equation}\label{distance}
z = \sqrt{\left( \Delta h \right)^{2}+2\Delta h R + R^{2}\cos^{2}\theta} - R\cos\theta,
\end{equation}
where we have defined
\begin{eqnarray}\label{altitude}
\nonumber \Delta h &=& \sqrt{R^{2} + z^{2} + 2z R\cos\theta} - R, \\
R &=& R_{\text{E}} + h_{0}.
\end{eqnarray}
In this manuscript, we have considered zenith communication, i.e. $\theta=0$.

\begin{figure}[t]
{\includegraphics[width=0.45 \textwidth]{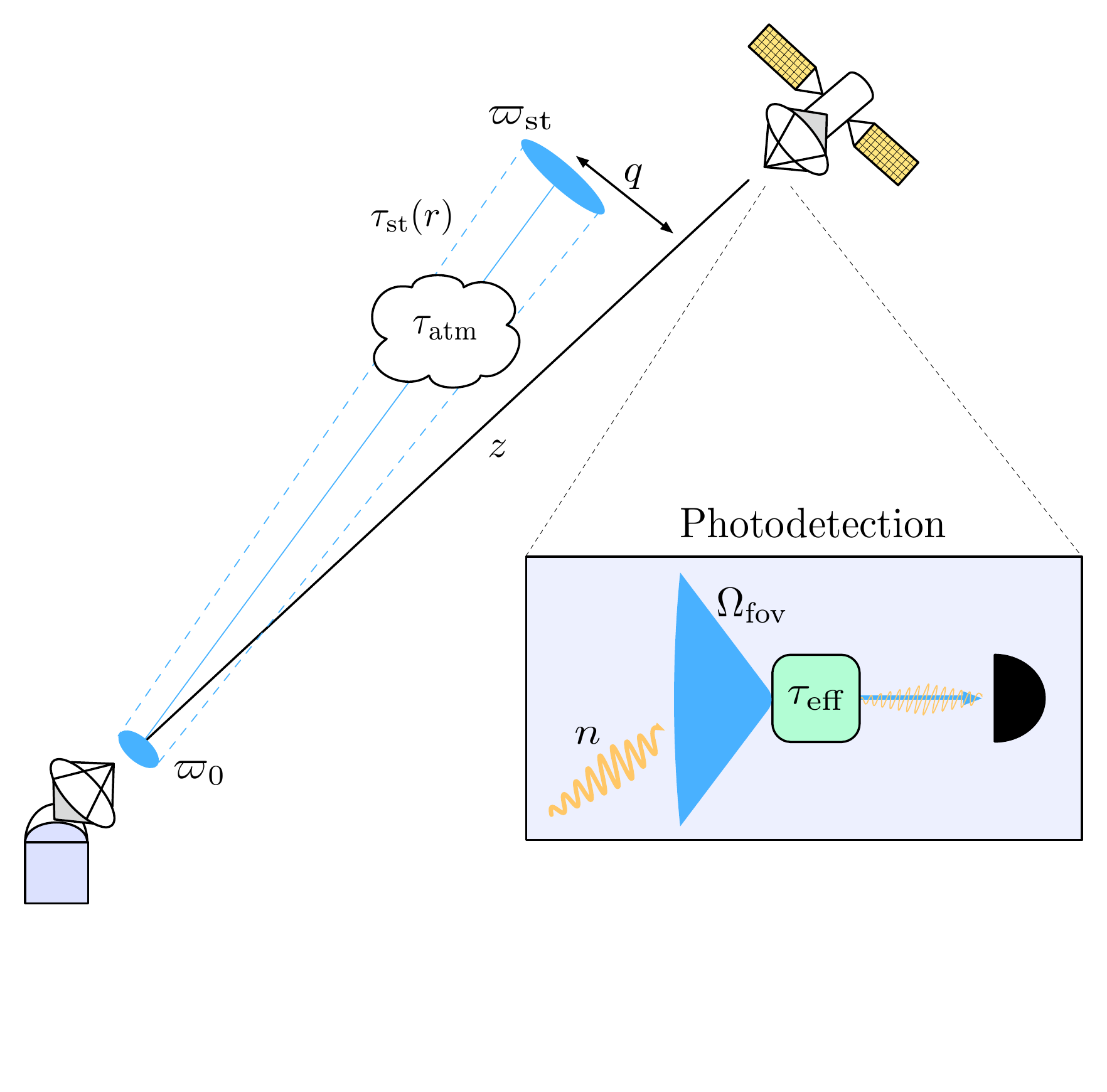}}
\caption{Quantum communication channel between a ground station and a satellite. A Gaussian beam (shown in blue) is generated at the ground with an initial waist $\varpi_{0}$, and propagates a distance $z$ through free space, where it suffers from diffraction and turbulence effects, as well as atmospheric absorption. These mechanisms induce transmissivities $\tau_{\text{st}}$ and $\tau_{\text{atm}}$, respectively. Apart from the broadening of the beam waist, $\varpi_{\text{st}}$, caused by turbulence, we also have wandering of the beam centroid, quantified by the distance $q$. The efficiency of the photodetectors is represented by the transmissivity $\tau_{\text{eff}}$. Given the field of view, $\Omega_{\text{fov}}$, there is a mean number of thermal photons $n$ detected by the receiver.}
\label{fig2}
\end{figure}

In order to understand the limitations of entanglement distribution and quantum teleportation in free space, we need a comprehensive study of loss mechanisms. We will describe them through attenuation channels with transmissivity $\tau_{i}$, which act on the modes of a given quantum state as
\begin{equation}
\hat{a} \longrightarrow \sqrt{\tau_{i}}\hat{a} + \sqrt{1-\tau_{i}}\hat{a}_{\text{th}}.
\end{equation}
Here, we will consider that these attenuation channels incorporate a thermal mode from the environment, represented here by $\hat{a}_{\text{th}}$. If we assume that the quantum state is propagating through a homogeneous thermal environment, then the composite effect of $N$ attenuation channels is represented by the action of a single one whose effective transmissivity is $\tau = \prod_{i=1}^{N}\tau_{i}$. 

We sketch the general quantum communication scenario in Fig.~\ref{fig2}. We will consider the combined effects of different loss mechanisms that apply to signals in the optical regime propagating through free space. These mechanisms have been identified in previous works studying quantum communication links involving ground stations and satellites~\cite{Vasylyev2019,Pirandola2021_2,Pirandola2021}.

\subsection{Diffraction}
We consider the effects of diffraction in signals propagating through free space. We assume a quasi-monochromatic bosonic mode represented by a Gaussian beam with wavelength $\lambda$, curvature radius of the wave-front $R_{0}$, and initial waist $\varpi_{0}$. For a focused beam, $R_{0}$ is equal to the distance between transmitter and receiver, whereas for a collimated beam, it is set at infinity. The receiver aperture is $a_{R}$, and $z_{R}=\pi\varpi_{0}^{2}/\lambda$ is the Rayleigh range, such that the far-field regime is defined by $z\gg z_{R}$, for a transmission distance $z$.

The transmissivity induced by diffraction is given by
\begin{equation}\label{tau_diffraction}
\tau_{\text{diff}} = 1 - e^{-2\left(a_{R}/\varpi_{z}\right)^{2}},
\end{equation}
where $\varpi_{z}$ is the waist of the beam at a distance $z$~\cite{Svelto2010}, 
\begin{equation}
\varpi_{z}^{2} = \varpi_{0}^{2}\left[ \left( 1-\frac{z}{R_{0}}\right)^{2} + \left( \frac{z}{z_{R}}\right)^{2}\right].
\end{equation}
Here we will work with collimated beams, for which we have
\begin{equation}
\varpi_{z}^{2} = \varpi_{0}^{2}\left[ 1 + \left( \frac{z}{z_{R}}\right)^{2}\right].
\end{equation}
Notice that losses associated to diffraction will be larger when $a_{R}\ll\varpi_{z}$.

\subsection{Atmospheric attenuation}
The transmissivity affected by atmospheric attenuation of signals at a fixed altitude is given by
\begin{equation}
\tau_{\text{atm}} = \text{exp}\left[-\alpha_{0}z e^{-h/\tilde{h}}\right],
\end{equation}
where $\alpha_{0} = N_{0}\sigma$ is the extinction factor, $N_{0}$ is the density of particles, $\sigma=\sigma_{\text{abs}}+\sigma_{\text{sca}}$ is the cross section associated with absorption and scattering, and $\tilde{h}=6600$ m is a scale factor~\cite{Vasylyev2019}. At sea level, and for $\lambda=800\text{ nm}$, we have $\alpha_{0}=5\times 10^{-6}\text{ m}^{-1}$. Naturally, this needs to be adapted to variable altitudes. For that, we will use Eqs.~\eqref{distance} and~\eqref{altitude}, considering $h_{0}$ to be negligible. Then, we can write
\begin{equation}
\tau_{\text{atm}} = e^{-\alpha_{0}g(h,\theta)},
\end{equation}
where we have defined
\begin{equation}
g(h,\theta) = \int_{0}^{z(h,\theta)} \text{d}y e^{-h(y,\theta)/\tilde{h}}.
\end{equation}

\subsection{Detector efficiency and thermal background}
As another source of losses, we can consider that we may have inefficient detectors. We will take, as the lowest value, $\tau_{\text{eff}}=0.4$~\cite{Liorni2019}, whereas the maximum possible one is $\tau_{\text{eff}}=1$. We will refer to the latter as the ideal case. Nevertheless, we consider that the signal traveling through the link will acquire excess noise that will be caught in the detectors, characterized by a thermal state that introduces $\tau_{\text{eff}}\,n$ thermal photons into our state. Furthermore, we consider that the effective number of thermal photons that the signal acquires in the path is the one that can be effectively captured by the detectors. We compute this using~\cite{Pirandola2021}
\begin{eqnarray}\label{thermal_background}
\nonumber \Gamma_{R} &=& \Delta\lambda \Delta t\Omega_{\text{fov}}a_{R}^{2}, \\
\nonumber N_{BB} &=& 2c\lambda^{-4}\left[ e^{hc/(\lambda k_{B}T)}-1\right]^{-1}, \\
n &=& \Gamma_{R}N_{BB},
\end{eqnarray}
where $\Gamma_{R}$ is the photon collection parameter, $\Delta t$ and $\Delta\lambda$ are the spectral filter and the time bandwidth, respectively, $\Omega_{\text{fov}}$ is the field of view of the receiver, and $N_{BB}$ is the number of thermal photons, quantified by the black-body formula, in units of $\Gamma_{R}^{-1}$. Furthermore, $c$ is the speed of light, $\lambda$ is the wavelength of the signal, $h$ is Planck's constant, $k_{B}$ is Boltzmann's constant, $T$ is the temperature, and finally $n$ is the average thermal photon number.

We take $\Delta\lambda=1$ nm, $\Delta t=10$ ns, $\Omega_{\text{fov}}=10^{-10}$ sr and $a_{R}=40$ cm. Then, we are left with average thermal-photon number $n_{\text{day}}^{\text{down}}=0.30$ and $n_{\text{night}}^{\text{down}}=3.40\times10^{-6}$ for a daytime and nighttime downlink, respectively, and $n_{\text{day}}^{\text{up}}=0.22$ and $n_{\text{night}}^{\text{up}}=5.43\times10^{-7}$ for a daytime and nighttime uplink, respectively.

\subsection{Turbulence}

Let us now look at the effects of turbulence. We aim at working in the weak-turbulence regime, in which the effects of scintillation are ignored. This regime can be characterized using the spherical-wave coherence length,
\begin{eqnarray}
\nonumber \rho_{0} &=& [1.46 k^{2}I_{0}(z)]^{-3/5}, \\
I_{0}(z) &=& \int_{0}^{z} \text{d}\xi \left( 1 - \frac{\xi}{z}\right)^{5/3}C^{2}_{n}(h(\xi,\theta))
\end{eqnarray}
through the following formula
\begin{equation}
z \lesssim k\left(\text{min}\{2a_{R},\rho_{0}\}\right)^{2},
\end{equation}
for a beam with wavenumber $k=2\pi/\lambda$, propagation distance $z$, and refraction index structure constant $C_{n}^{2}$. The latter, in the Hufnagel-Valley model of atmospheric turbulence~\cite{Hufnagel1964,Valley1980}, reads
\begin{eqnarray}
\nonumber C^{2}_{n} &=& 5.94\times10^{-53}\left( \frac{v}{27}\right)^{2}h^{10}e^{-h/1000} \\
&+& 2.7\times 10^{-16}e^{-h/1500} + A e^{-h/100},
\end{eqnarray}
and it measures the strength of the fluctuations in the refraction index caused by spatial variations of temperature and pressure. In this manuscript, we consider $v=21\text{ m}/\text{s}$ for the wind speed, and $A_{\text{day(night)}}=2.75 (1.7)\times10^{-14} \text{ m}^{-2/3}$ for daytime (nighttime) values. At constant altitude, we see that $\rho_{0}=(0.548 k^{2} C_{n}^{2}z)^{-3/5}$. If we consider an uplink, we can use the above formula, but for a downlink, we need to substitute $\xi\rightarrow z-\xi$ in the structure constant.

In the weak-turbulence regime, we can distinguish between two error sources, caused by the interaction of the beam with vortices, or \textit{eddies}, of different sizes: beam broadening and beam wandering~\cite{Fante1975,Fante1980}. Beam broadening is caused by eddies smaller than the beam waist, and acts on a fast time scale. This will replace $\varpi_{z}$ by some short-term waist $\varpi_{\text{st}}$, leading to the modified diffraction-induced transmissivity
\begin{equation}
\tau_{\text{st}} = 1 - e^{-2\left(a_{R}/\varpi_{\text{st}}\right)^{2}}.
\end{equation}
Here, we can write
\begin{equation}
\varpi_{\text{st}}^{2} \simeq \varpi_{z}^{2} + 2\left( \frac{\lambda z}{\pi\rho_{0}}\right)^{2}(1-\phi)^{2},
\end{equation}
where $\phi = 0.33(\rho_{0}/\varpi_{0})^{1/3}$. In the weak-turbulence regime, we find that $\phi\ll1$, and thus we can approximate $(1-\phi)^{2} \approx 1-0.66(\rho_{0}/\varpi_{0})^{1/3}$~\cite{Yura1973}. 

Beam wandering is caused by eddies larger than the beam waist, and act on a slow time scale. This causes the beam to deflect by randomly displacing its center, leading to a wandering of the waist. This random displacement will be assumed to follow a Gaussian probability distribution with variance $\sigma^{2}$, which will be composed of the large-scale turbulence $\sigma^{2}_{\text{TB}}$ and the pointing error $\sigma^{2}_{\text{P}}$ variances. The long-term waist of the beam can be approximated by
\begin{equation}
\varpi_{\text{lt}}^{2} \simeq \varpi_{z}^{2} + 2\left( \frac{\lambda z}{\pi\rho_{0}}\right)^{2},
\end{equation}
and it is related to its short-term counterpart through 
\begin{equation}
\sigma_{\text{TB}}^{2} = \varpi_{\text{lt}}^{2} - \varpi_{\text{st}}^{2} \simeq \frac{0.1337\lambda^{2}z^{2}}{\varpi_{0}^{1/3}\rho_{0}^{5/3}}.
\end{equation}
The beam centroid wanders with total variance $\sigma^{2}=\sigma_{\text{TB}}^{2}+\sigma_{\text{P}}^{2}$, and we will take $\sigma_{\text{P}}=10^{-6}z$. We define $q$ as the distance between the beam centroid and the original center (horizontally-aligned with the transmitter and the receiver), also known as deflection. Following Ref.~\cite{Vasylyev2012}, we assume that this value takes a Gaussian random walk following the Weibull distribution
\begin{equation}
P_{\text{WB}}(q) = \frac{q}{\sigma^{2}}e^{-\frac{q^{2}}{2\sigma^{2}}}.
\end{equation}
Then, the maximum value of the transmissivity occurs for $q=0$,
\begin{equation}
\tau_{\text{max}} = \tau(q=0) = \tau_{\text{st}}(q=0)\tau_{\text{atm}}\tau_{\text{eff}}.
\end{equation}
However, for each instantaneous value of $q$, there will be an instantaneous $\tau(q)\leq\tau_{\text{max}}$ happening with a probability $P(\tau)$. The transmissivity associated to diffraction modified by this behavior is then~\cite{Vasylyev2012, Pirandola2021}
\begin{equation}
\tau_{\text{st}}(q) = e^{-4(q/\varpi_{\text{st}})^{2}}Q_{0}\left(\frac{2s^{2}}{\varpi_{\text{st}}^{2}}, \frac{4 q a_{R}}{\varpi_{\text{st}}^{2}}\right),
\end{equation}
where $Q_{0}(x,y)=\frac{e^{x}}{2x}\int_{0}^{y}\text{d}t \, te^{-t^{2}/4x}I_{0}(t)$ is an incomplete Weber integral and $I_{n}$ is the modified Bessel function (of order $n$) of the first kind. We can express
\begin{equation}
\tau(q) = \tau_{\text{max}}e^{-\left(q/q_{0}\right)^{\gamma}},
\end{equation}
where we have defined
\begin{eqnarray}
\nonumber \tau_{\text{st}}^{\text{far}} &=& \frac{2a_{R}^{2}}{\varpi_{\text{st}}^{2}}, \\
\nonumber \Lambda_{n}(x) &=& e^{-2x}I_{n}(2x), \\
\nonumber \gamma &=& \frac{4\tau_{\text{st}}^{\text{far}}\Lambda_{1}\left(\tau_{\text{st}}^{\text{far}}\right)}{1-\Lambda_{0}\left(\tau_{\text{st}}^{\text{far}}\right)}\left[ \log\left( \frac{2\tau_{\text{st}}}{1-\Lambda_{0}\left(\tau_{\text{st}}^{\text{far}}\right)}\right)\right]^{-1}, \\
q_{0} &=& a_{R}\left[ \log\left( \frac{2\tau_{\text{st}}}{1-\Lambda_{0}\left(\tau_{\text{st}}^{\text{far}}\right)}\right)\right]^{-1/\gamma}.
\end{eqnarray}
The probability distribution over $q$ induces then another probability distribution over $\tau$, 
\begin{equation}
\nonumber P(\tau) = \frac{q_{0}^{2}}{\gamma\sigma^{2}\tau}\left( \log\frac{\tau_{\text{max}}}{\tau}\right)^{\frac{2}{\gamma}-1}\exp\left[ -\frac{q_{0}^{2}}{2\sigma^{2}}\left( \log\frac{\tau_{\text{max}}}{\tau}\right)^{\frac{2}{\gamma}}\right].
\end{equation}
This function can be obtained from the Weibull distribution $P_{\text{WB}}(q)$ by using 
\begin{equation}
P(\tau) = \left. p(q|\sigma)\right|_{q=q(\tau)} \left| \frac{\text{d}q}{\text{d}\tau}\right|
\end{equation}
together with
\begin{equation}
q = q_{0}\left( \log\frac{\tau_{\text{max}}}{\tau}\right)^{\frac{1}{\gamma}}.
\end{equation}

\section{Entanglement distribution and quantum teleportation}
The quantum channel, once characterized by transmissivity $\tau$, is now described by the ensemble $\mathcal{E}=\{\mathcal{E}_{\tau},P(\tau)\}$, where the channel $\mathcal{E}_{\tau}$ is selected at random with a probability taken from the distribution $P(\tau)$. This is called a fading channel.


We will use this to describe the degradation of entanglement on states propagating through free space, which we will quantify through the negativity of the covariance matrix of Gaussian states, and through the average fidelity of teleporting an unknown coherent state using the entangled resources. We will consider two-mode squeezed states as a typical case of bipartite CV entangled states. 

Since two-mode squeezed states are Gaussian, and the fading channel is Gaussian-preserving, we can use the covariance-matrix formalism to describe the evolution of the state. This will provide the obvious advantages of using finite-dimensional matrices to work with infinite-dimensional operators, but it will also lead to a convenient description of fading channels. Consider a two-mode Gaussian state with vanishing first moments and covariance matrix in normal form given by
\begin{equation}\label{CM}
\Sigma = \begin{pmatrix} \alpha\mathbb{1}_{2} & \gamma Z \\ \gamma Z & \beta\mathbb{1}_{2} \end{pmatrix},
\end{equation}
and consider a single-mode environment described by a Gaussian state with covariance matrix $E=m\mathbb{1}_{2}$. For example, for a daytime downlink, this state is characterized by $m=1+2\tau_{\text{eff}}n_{\text{day}}^{\text{down}}$. We assume that the second mode is the one transmitted through open air, and therefore it is affected by the fading channel. Keeping only the transmitted contribution, we obtain
\begin{equation}\label{CM_asym}
\Sigma' = \begin{pmatrix} \alpha\mathbb{1}_{2} & \left\langle\sqrt{\tau}\right\rangle\gamma Z \\ \left\langle\sqrt{\tau}\right\rangle\gamma Z & \left[ \langle\tau\rangle\beta + \left(1-\langle\tau\rangle\right)m \right] \mathbb{1}_{2} \end{pmatrix}.
\end{equation}
See Appendix~\ref{app_A} for more on Gaussian fading channels. This description assumes that turbulence is a fast process, or at least it is much faster than the detectors. Therefore we observe only an average characterization of the channel through $\langle\tau\rangle$. If we considered that the detectors were much faster than the turbulence, then we would obtain $\tau$ instead of $\langle\tau\rangle$, and we would have to average the obtained quantity afterwards. In this scenario, the quantum teleportation fidelity would be
\begin{equation}
\overline{F} = \int_{0}^{\tau_{\text{max}}}d\tau P_{0}(\tau)\overline{F}(\tau),
\end{equation}
which we will refer to as slow turbulence. In contrast, the teleportation fidelity in the fast turbulence regime is $\overline{F}\left(\langle\tau\rangle\right)$.

For an entangled Gaussian resource that has the covariance matrix in Eq.~\eqref{CM}, the average fidelity of teleporting an unknown coherent state is $\overline{F}=\left[ 1+\frac{1}{2}(\alpha+\beta-2\gamma) \right]^{-1}$. Now, if we introduce the effect of the fast fading channel, we see that
\begin{equation}
\frac{1}{\overline{F}} = \left\{ 1 + \frac{1}{2}\left[ \alpha + \langle\tau\rangle\beta + \left( 1 - \langle\tau\rangle\right)m -2\langle\sqrt{\tau}\rangle\gamma \right] \right\},
\end{equation}
while for the slow fading channel, the average is computed numerically. The other quantity in which we are interested is the negativity of the covariance matrix, a measure for entanglement. For two-mode Gaussian states, it is given by~\cite{Serafini2004} 
\begin{equation}
\mathcal{N} = \text{max}\left\{ 0, \frac{1-\tilde{\nu}_{-}}{2\tilde{\nu}_{-}}\right\},
\end{equation}
where $\tilde{\nu}_{-}$ is the smallest symplectic eigenvalue of the partially-transposed covariance matrix. For the one in Eq.~\eqref{CM}, it can be written as
\begin{equation}
\tilde{\nu}_{-} = \frac{\alpha+\beta-\sqrt{(\alpha-\beta)^{2}+4\gamma^{2}}}{2}.
\end{equation}
Notice that the condition for entanglement is $\tilde{\nu}_{-}<1$, which can be expressed as $(\alpha-1)(\beta-1)<\gamma^{2}$. 

Both the teleportation fidelity and the negativity are reduced because the entanglement of the state degrades while propagating through free space. The degradation is more severe with increasing distance, as the transmissivity of the fading channel decreases. Here, we investigate the teleportation fidelity and the negativity associated with a two-mode squeezed state with covariance matrix
\begin{equation}\label{TMSV}
\Sigma = \begin{pmatrix} \cosh2r\mathbb{1}_{2} & \sinh2r Z \\ \sinh2r Z & \cosh2r\mathbb{1}_{2} \end{pmatrix},
\end{equation}
where $r$ is the squeezing parameter, and it is directly related with the (initial) negativity through $\tilde{\nu}_{-}=e^{-2r}$, meaning no entanglement for $r=0$, and infinite entanglement for $r\rightarrow\infty$. The teleportation fidelity associated with using a TMSV state is $\overline{F} = \left( 1+e^{-2r} \right)^{-1}$~\cite{Pirandola2006}, and it reaches the maximum classical fidelity of $1/2$ for no squeezing ($r=0$), while approaching 1 for infinite squeezing ($r\rightarrow\infty$). 

\begin{figure*}[t]
{\includegraphics[width=0.95 \textwidth]{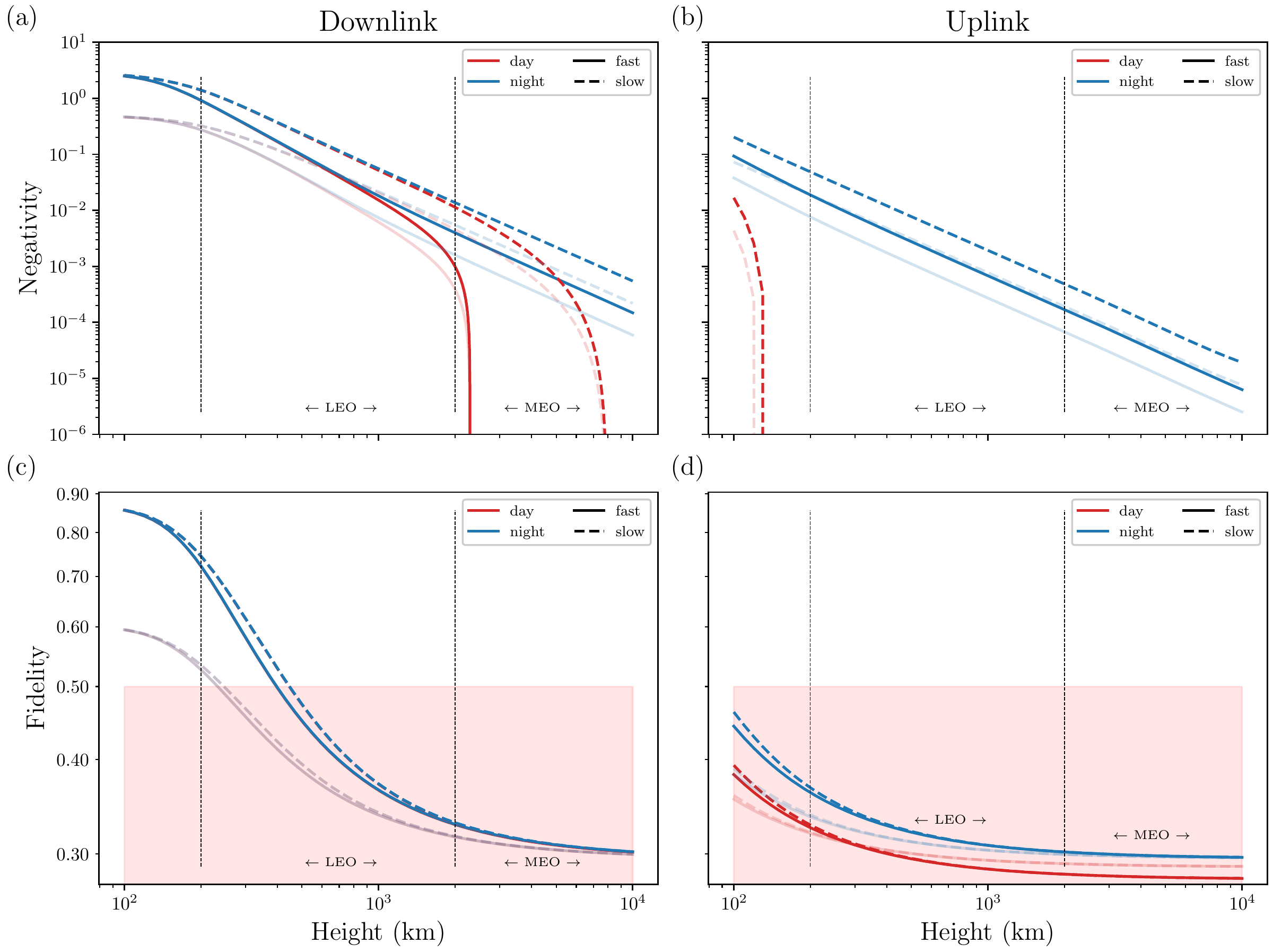}}
\caption{Quantum communication using a TMSV state distributed through free space, which has undergone a loss mechanism comprising diffraction, atmospheric extinction, detector inefficiency and free-space turbulence, for a signal with wavelength $\lambda=800$ nm, squeezing parameter $r=1$, and waist $\varpi_{0}=20$ cm, assuming the receiver has an antenna with aperture $a_{R}=40$ cm. We represent the negativity for (a) a downlink and (b) an uplink. We also represent the fidelity of quantum teleportation for coherent states using this entangled resource, for (c) a downlink and (d) an uplink. Dashed lines represent the regime of slow turbulence, and solid lines represent the regime of fast turbulence, when comparing them to the velocity of the detectors. Nighttime and daytime thermal noise is taken into account in the blue and red curves, respectively. In the case of the downlink fidelity, note that the results which incorporate daytime (red) and nighttime (blue) thermal noise coincide. In full color, we present the results for perfect detector efficiency, $\tau_{\text{eff}}=1$, whereas the high-transparency curves correspond to $\tau_{\text{eff}}=0.4$.}
\label{fig3}
\end{figure*}

In Figs.~\ref{fig3}~(a),~(b), we represent the negativity of a TMSV state with initial squeezing $r=1$ against the altitude of the link. Fig.~\ref{fig3}~(a) shows the results for a downlink, and Fig.~\ref{fig3}~(b) illustrates an uplink. In solid lines, we can see the results of a fast-turbulence scenario, whereas the dashed lines represent a slow-turbulence one. Furthermore, blue and red lines incorporate nighttime and daytime thermal noise, respectively. In full color, we can see the values associated with perfect detector efficiency, $\tau_{\text{eff}}=1$, whereas the lines with high transparency correspond to faulty detectors with $\tau_{\text{eff}}=0.4$. We can observe that the negativity is reduced exponentially with the distance, and we see better results for a downlink than for an uplink. In vertical lines, we mark zones associated to different orbital altitudes. These are the low-Earth orbit (LEO), from 200 km to 2000 km and the medium-Earth orbit (MEO), from 2000 km to 42164 km. Orbits from 42164 km on are known as geostationary orbits. 


Figs.~\ref{fig3}~(c),~(d) show the fidelity of a quantum teleportation protocol for coherent states, that uses TMSV states distributed through (c) a downlink or (d) an uplink through free space. The degradation of the entanglement of this state is due to the various loss mechanisms that comprise the fading channel: diffraction, atmospheric attenuation, detector inefficiency and turbulence. This degradation is responsible for the deterioration of the teleportation fidelity, which depends only on the entangled resource that is consumed. The slow-turbulence regime is represented by dashed lines, while the fast-turbulence regime is represented by solid lines. The red ones incorporate daytime thermal noise, whereas the blue ones consider nighttime thermal noise. Perfect detector efficiency ($\tau_{\text{eff}}=1$) is represented by full-color lines, while an imperfect detector ($\tau_{\text{eff}}=0.4$) was considered in the high-transparency lines. Here, we observe that only quantum teleportation protocols through a downlink in the LEO region can produce fidelities above the maximum classical result~\cite{Braunstein2001}; all instances worse than this are enclosed in a pale red background. Notice that, in Fig.~\ref{fig3}~(c), results for daytime and nighttime thermal noise coincide, both in the perfect and imperfect detector scenarios. This also happens for short distances in Fig.~\ref{fig3}~(a).


\subsection{Intermediate station for state generation}
We have observed that the effects of turbulence are more severe in the atmosphere and have stronger effects on signals that have not suffered diffraction. Therefore, the scenario in which we have an uplink path presents more difficulties for free-space entanglement distribution. Nevertheless, we investigate a scenario in which there is an intermediate station connecting the ground station and the satellite, and we consider that TMSV states can be generated at this intermediate station. Our goal is to observe whether there is an increase in the entanglement available when the distance that the signals travel through free space is reduced. This already presents an advantage, because now the uplink does not start at the Earth, but at a given orbit, and the turbulence effects are highly reduced. 

\begin{figure*}[t]
{\includegraphics[width=0.95 \textwidth]{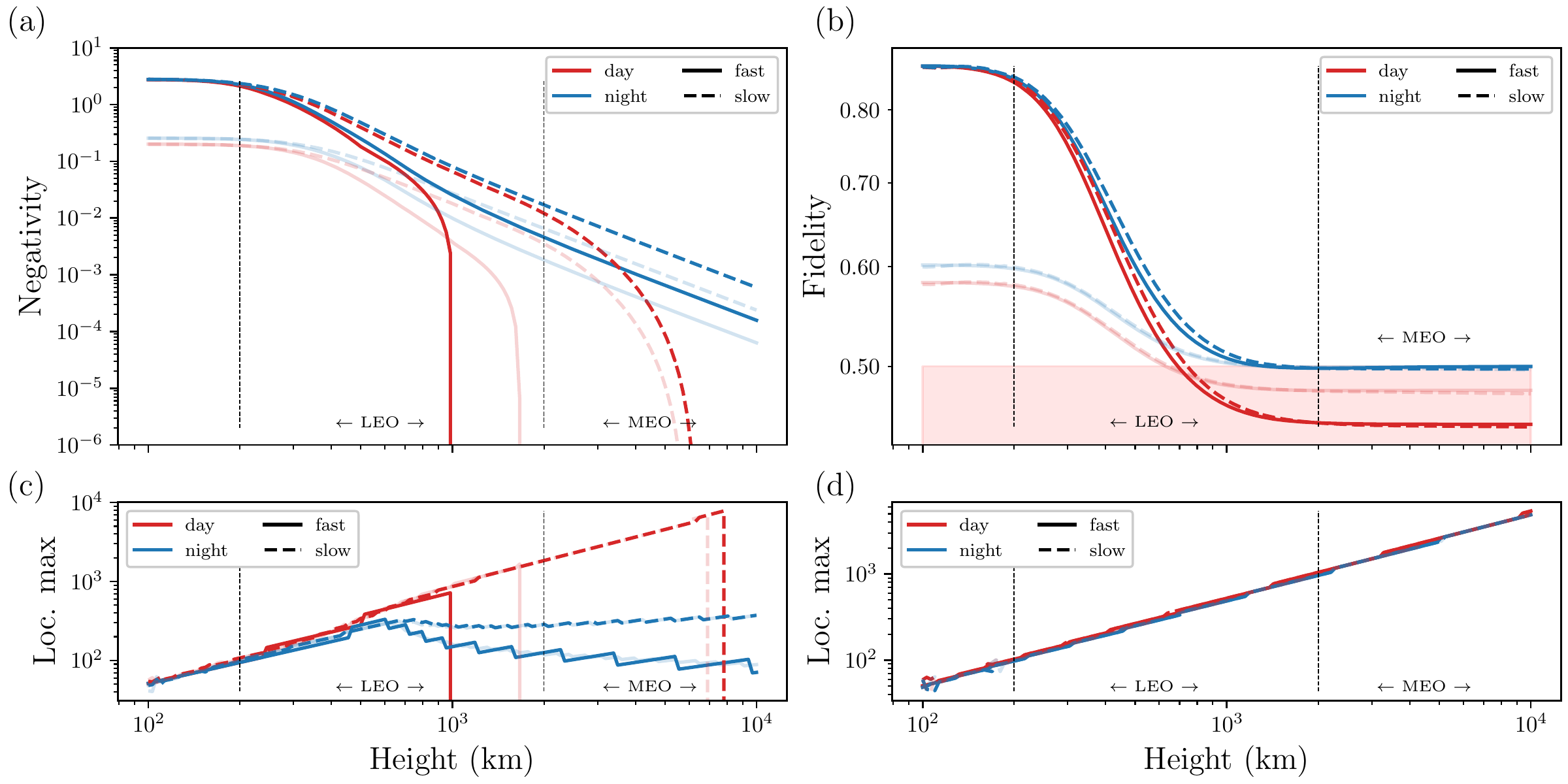}}
\caption{(a) Negativity of transmitted TMSV states in free-space communication against the height of the satellite link, with respect to the ground station. We consider that these stated are generated in an intermediate station, where one mode is sent to the ground station and the other to the satellite. We study a signal with wavelength $\lambda=800$ nm, squeezing parameter $r=1$, and initial waist $\varpi_{0}=20$ cm, sent to a receiver that has an antenna of radius $a_{R}=40$ cm. This signal is subject to a loss mechanism composed of diffraction, atmospheric extinction, detector inefficiency and free-space turbulence, described by a fading channel. In this case, the results that incorporate daytime (red) and nighttime (blue) thermal noise coincide. We distinguish between the results obtained in the slow-turbulence and fast-detection regime, in dashed lines, and the fast-turbulence and slow-detection regime, in solid lines. The results for perfect detector efficiency, $\tau_{\text{eff}}=1$, appear in full color, whereas the transparent curves correspond to $\tau_{\text{eff}}=0.4$. (b) Fidelity of teleporting an unknown coherent state using these entangled resources. We represent the optimal position of the intermediate station, for the different turbulence conditions, that achieve the maximum possible negativities in (c), and those that lead to maximum fidelities, in (d).}
\label{fig4}
\end{figure*}

In this case, the covariance matrix of the two-mode Gaussian state, after a single application of the fading channel, is
\begin{equation}\label{CM_sym}
\Sigma' = \begin{pmatrix} \left[\tau_{\text{d}}\alpha+(1-\tau_{\text{d}})m_{\text{d}}\right]\mathbb{1}_{2} & \sqrt{\tau_{\text{d}}\tau_{\text{u}}}\gamma Z \\ \sqrt{\tau_{\text{d}}\tau_{\text{u}}}\gamma Z & \left[ \tau_{\text{u}}\beta + \left(1-\tau_{\text{u}}\right)m_{\text{u}} \right] \mathbb{1}_{2} \end{pmatrix},
\end{equation} 
where we define by $\tau_{\text{d(u)}}$ the transmissivity of the fading channel describing signal propagation through the downlink (uplink). After multiple applications of the fading channel, in the case of fast turbulence and slow detection, we will have that the negativity and the teleportation fidelity can be averaged as $\mathcal{N}=\mathcal{N}\left(\langle\tau_{\text{d}}\rangle,\langle\tau_{\text{u}}\rangle\right)$ and $\overline{F} = \overline{F}\left(\langle\tau_{\text{d}}\rangle,\langle\tau_{\text{u}}\rangle\right)$, respectively. On the opposite regime, slow turbulence and fast detection, these averages are computed as
\begin{eqnarray}
\nonumber \mathcal{N} &=& \int_{0}^{\tau_{\text{d}^{\text{max}}}} \text{d}\tau_{\text{d}}P(\tau_{\text{d}}) \int_{0}^{\tau_{\text{u}^{\text{max}}}} \text{d}\tau_{\text{u}} P(\tau_{\text{u}}) \mathcal{N}\left(\tau_{\text{d}},\tau_{\text{u}}\right), \\
\overline{F} &=& \int_{0}^{\tau_{\text{d}^{\text{max}}}} \text{d}\tau_{\text{d}}P(\tau_{\text{d}}) \int_{0}^{\tau_{\text{u}^{\text{max}}}} \text{d}\tau_{\text{u}} P(\tau_{\text{u}}) \overline{F}\left(\tau_{\text{d}},\tau_{\text{u}}\right).
\end{eqnarray}

\begin{figure*}[t]
{\includegraphics[width=0.95 \textwidth]{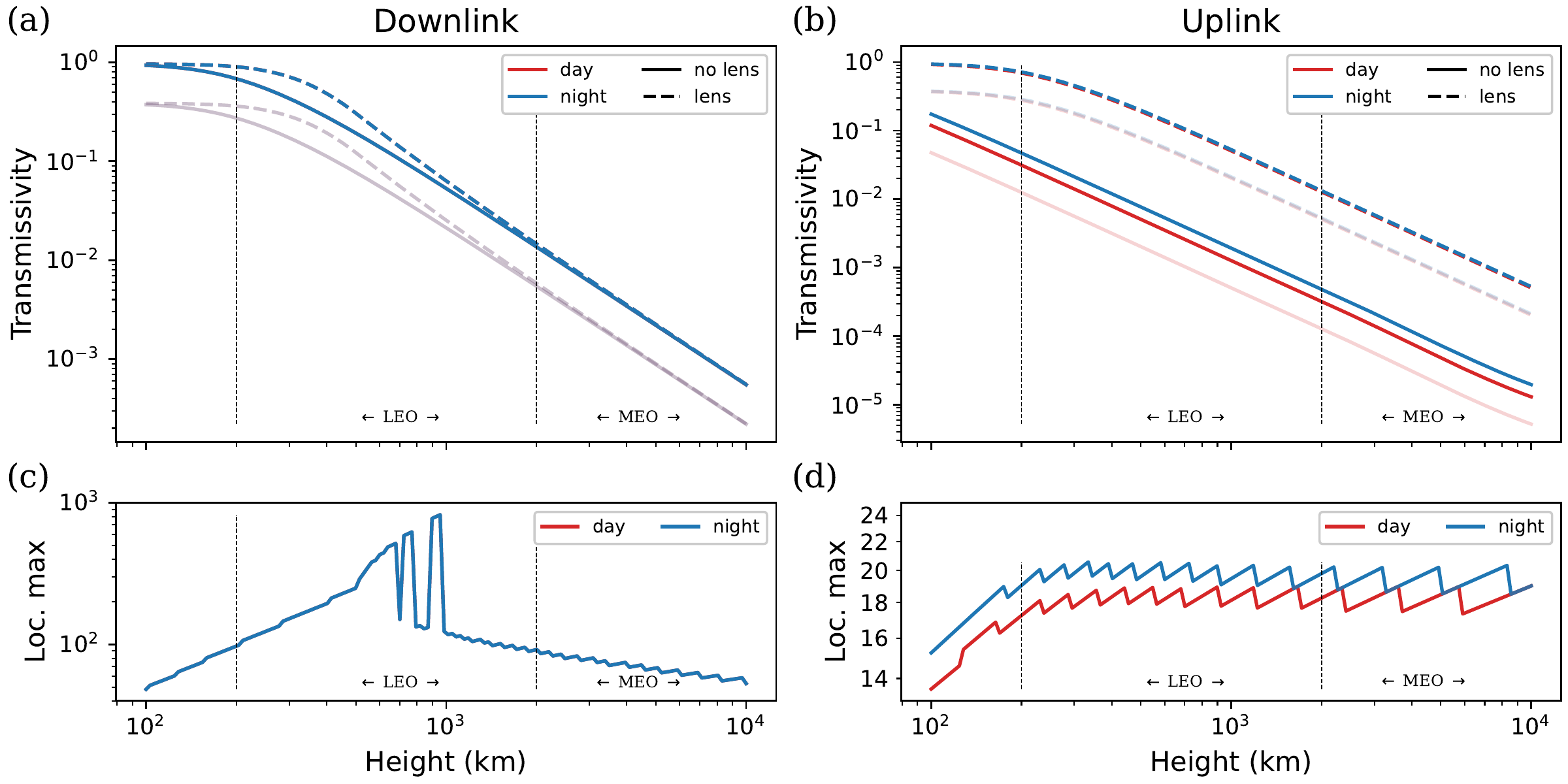}}
\caption{Average transmissivity of a fading channel modeling a loss mechanism present in a link connecting a ground station and a satellite, and composed of diffraction, atmospheric extinction, detector inefficiency and free-space turbulence. In a solid line, we plot the quantities associated an unaltered channel, and in a dashed line, we represent the cases in which a lens has been placed in a mid point of the link to reduce beam broadening. In full color we represent the results for perfect detector efficiency, $\tau_{\text{eff}}=1$, whereas imperfect detection, $\tau_{\text{eff}}=0.4$, is marked by the transparent curves. In blue, we represent the result associated with nighttime thermal noise, whereas those associated to daytime thermal noise appear in red. We represent the results associated to a downlink in (a), with the optimal location of the lens, in order to maximize the transmissivity, given in (c). The results for nighttime thermal noise fall on top of those for daytime thermal noise. The results associated to an uplink are represented in (b), with optimal positions of the lens shown in (d).}
\label{fig5}
\end{figure*}

In Fig.~\ref{fig4}~(a), we represent the negativity of the final state, considering an optimal placement of the intermediate station, for each value of the total height. These optimal points are shown in Fig.~\ref{fig4}~(c). We can observe that the results are improved, with respect to both the downlink and the uplink. We can also observe this improvement, especially with respect to the uplink, and remarkably for high altitudes, in Fig.~\ref{fig4}~(b). Here, we represent the fidelity of teleporting an unknown coherent state using TMSV states, generated at the intermediate station, and having both modes distributed through the noisy and turbulent links. Only the fidelities with nighttime thermal noise remain above the maximum classical fidelity of $1/2$, while the accumulated thermal noise in daytime links leads to fidelities that fall below this limit at altitudes in the LEO region. We can see that the limit is extended with respect to the downlink, and the fidelity for an uplink never achieved values above it. Therefore, the generation of entangled states in an intermediate station between the ground station and the satellite greatly improves the teleportation fidelity. 

In the case of the negativity with an intermediate station, we observe an improvement, especially in the case of ideal detectors; for imperfect ones, represented by $\tau_{\text{eff}}=0.4$, the results do not differ significantly from those of the downlink. This is because, for an intermediate station, we are considering now two detection events, instead of one, which enhances the error in the case of imperfect detectors. These comparisons are illustrated in Appendix~\ref{app_C} (see Figs.~\ref{fig8}~(a) and (b) there). On the contrary, the results for the teleportation fidelity are highly improved with an intermediate station, and extend also to the case of imperfect detectors ({\color{red} cf.} Figs.~\ref{fig8}~(c) and (d) in Appendix~\ref{app_C}). Although, for imperfect detectors , fidelities with daytime thermal noise can go below the maximum classical fidelity. Of course, it would be natural to assume that we obtain good results for the fidelity because we are optimizing the placement of the intermediate station and keeping the highest fidelity at each altitude. And rightly so, but the improvement difference in the negativity and the fidelity is due to the fact that, when considering the intermediate station, the states generated there, and distributed through a downlink to Earth and though an uplink to a satellite, which makes them more symmetric. In the case of a single downlink or uplink, one of the modes was kept and the other was sent through free space, resulting in a covariance matrix that was highly asymmetric (see Eq.~\eqref{CM_asym}). Given two Gaussian quantum states with the same negativity, the one whose covariance matrix is more symmetric shows higher teleportation fidelity. For further discussion, see Appendix~\ref{app_C}. 

\subsection{Intermediate station for beam focusing}
Here, we consider using the intermediate station as a point where the signal is refocused, in an attempt to reduce the effects of diffraction and turbulence. This could improve the transmissivity of the downlink, but especially that of the uplink, where the turbulence effects are more damaging.  This is what we observe in Fig.~\ref{fig5}, where we represent the transmissivity of the fading channel describing the propagation through the link, against the total height. In Fig.~\ref{fig5}~(a) the transmissivity for a downlink is improved in the ideal case, similarly to how it was improved by generating the states in the intermediate station; in this case, however, we consider that the sender generates both modes, and thus only have one detector at the receiver. Furthermore, notice that in Fig.~\ref{fig5}~(c) the optimal location of the intermediate lens is very similar to the optimal position of the intermediate station in Appendix~\ref{app_C} (see Fig.~\ref{fig9}~(b)). This emphasizes the statement that an intermediate station and an intermediate lens contribute about equally to improving the transmissivity of the channel, considering a downlink. However, when we see the case of an uplink in Fig.~\ref{fig5}~(b), we notice that it is substantially improved, achieving values above the transmissivity of the downlink. This is because the optimal locations of the focusing lens, represented in Fig.~\ref{fig5}~(d), all fall in a range of tens of kilometres, very close to the ground station, in order to reduce the effects of turbulence inside the atmosphere. As a last remark, see that the results the results for daytime and nighttime thermal noise coincide for certain ranges, in Fig.~\ref{fig5}, both in downlink and uplink scenarios. 

\section{Microwave slant links}
We aim at expanding the results shown in this manuscript by considering the attenuation of microwave quantum signals in free-space propagation. The major difference with the model for signals in the optical regime will be the omission of turbulence effects. Given the wavelengths for microwaves, on the order of centimetres, we can see that they will not be affected by the fluctuations that lead to turbulence for optical signals. Nevertheless, also because of the long wavelengths, microwaves will be highly affected by diffraction. By proposing a loss mechanism composed of diffraction, atmospheric attenuation and detector inefficiency, we aim at investigating the limits for entanglement distribution and quantum teleportation with microwaves in free space. With the diffraction-induced transmissivity given in Eq.~\eqref{tau_diffraction}, and assuming ideal detector efficiency $\tau_{\text{eff}}=1$, we describe the absorption-induced transmissivity along a slant path of zenit angle $\theta$, starting at altitude $h_{0}$ and ending at $h$ by
\begin{equation}
\tau_{\text{atm}} = \exp\left[ -\text{sec}\theta\int_{h_{0}}^{h}\text{d}h'\alpha(h')\right].
\end{equation}
the atmospheric absorption coefficient $\alpha(h')=\alpha_{o}+\alpha_{w}(h')$ represents the combined attenuation due to oxygen and water vapor. The former can be considered constant inside the atmosphere, but the latter will depend on the variation of the water concentration with the altitude. The specific coefficients are~\cite{Ho2004}
\begin{eqnarray}
\nonumber \alpha_{o} &=& 1.44\times10^{-3} \text{ km}^{-1}, \\
\nonumber \alpha_{w}(h') &=& 4.44\times10^{-5} p_{0}e^{-\frac{h'}{2}} \text{ km}^{-1}, 
\end{eqnarray}
where $p_{0}$ is the water-vapor density, whose average ground value is $7.5 \text{ g/m}^{3}$, at 5 GHz. These frequencies present one of the lowest attenuation profiles among microwaves~\cite{ITU-R}, and therefore make them suitable for telecommunications independent of the weather conditions. However, the main sources of loss for microwave signals are diffraction and the thermal background. 

Due to the bright thermal background that microwave present at room temperatures, these states are generated at cryogenic temperatures; nevertheless, we consider that the squeezing operations are applied to a thermal state, and not to an ideal vacuum state, which leads to the more realistic  two-mode squeezed thermal state. This is also a Gaussian state, with covariance matrix
\begin{equation}\label{TMST}
\Sigma = (1+2n)\begin{pmatrix} \cosh2r\mathbb{1}_{2} & \sinh2r Z \\ \sinh2r Z & \cosh2r\mathbb{1}_{2} \end{pmatrix}.
\end{equation}
Our choice of entangled resource describes a two-mode squeezed thermal (TMST) state, characterized by $n = 10^{-2}$ average number of thermal photons per mode, and squeezing parameter $r=1$. 

In order for these states to remain entangled when distributed through free space, we need the transmissivity of the channel to satisfy
\begin{equation}
\tau > \frac{(m-1)(c-1)}{(m-c)(c-1)+s^{2}},
\end{equation}
assuming that $m>c$, for a state represented by the covariance matrix in Eq.~\eqref{CM_asym} with $\alpha=\beta=c$ and $\gamma=s$. If the state is symmetric, and its covariance matrix resembles that in Eq.~\eqref{CM_sym}, with $\tau_{\text{d}}\approx\tau_{\text{u}}\equiv\tau$, this condition turns to
\begin{equation}
\tau > \frac{m-1}{m-c+s}.
\end{equation}
Considering identical initial resources (see Eq.~\eqref{TMSV}), this condition is always more restrictive for symmetric ($\tau>0.9997$) than for asymmetric states ($\tau>0.9992$, given the states studied here).

We can reduce the effects of thermal noise if we assume that we know the time of arrival of the signal, and therefore by using Eq.~\eqref{thermal_background}. In Ref.~\cite{Pirandola2021_3}, the limits for short-range microwave QKD were studied, using as parameters $\Delta\, t \, \Delta \, \nu\simeq 1$. This lead to $\Gamma_{R}\simeq\lambda^{2}\Omega_{\text{fov}}a_{R}^{2}/c$ and, by taking $\lambda=6$ cm, $\Omega_{\text{fov}}=10^{-4}$ sr and $a_{R}=2$ m, the number effective number of thermal photons becomes $n\simeq 266$ at 288 K. 

\begin{figure}[t]
{\includegraphics[width=0.48 \textwidth]{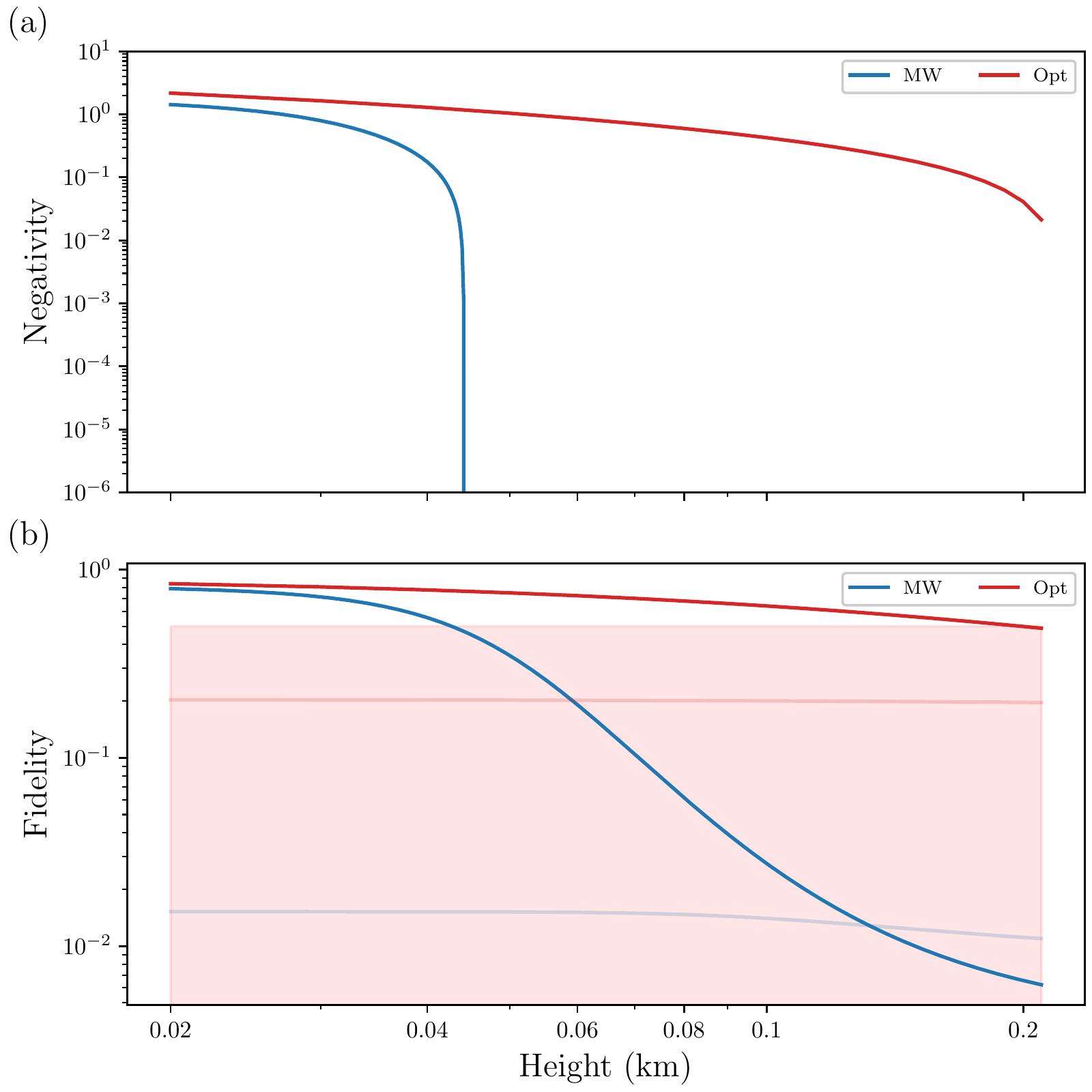}}
\caption{Performance comparison between microwave and optical signals in free space quantum communications under severe weather conditions. We represent the negativity (a) and the quantum teleportation fidelity (b) for TMSV states generated at a ground station at an altitude of 10 m, and where one of the modes is sent through an uplink. We consider the signal has squeezing parameter $r=1$, and initial waist $\varpi_{0}=1$ m, assuming the receiver has an antenna of radius $a_{R}=2$ m. In red, we represent the results associated to a signal in the optical regime, with wavelength $\lambda=800$ nm and zero thermal photons, whereas in blue we represent the results for microwave signals with wavelength $\lambda=6$ cm and $n=10^{-2}$ thermal photons. For optical signals, the thermal noise coming from the environment is characterized by $13.57$ photons, whereas for microwave signals, we have $266$ photons. In full color, we present the results for perfect detector efficiency, $\tau_{\text{eff}}=1$, whereas the transparent curves correspond to $\tau_{\text{eff}}=0.4$. The pale red background represents the region in which the teleportation fidelity falls below the maximum classical value.}
\label{fig6}
\end{figure}

With this, the condition for entanglement preservation on asymmetric states becomes $\tau>0.996$. Then, we see that the entanglement-distribution limit is 44 m, while the fidelity reaches the classical limit at 43 m. In this case, the asymmetry between both modes of the state distributed through free space does not lead to a significant difference between entanglement preservation and quantum teleportation distances. For symmetric states, the condition for symmetric states is $\tau>0.998$. Considering an intermediate station for state generation, the entanglement-distribution and quantum teleportation limit extends to 49 m. On the other hand, an intermediate station for beam refocusing leads to a limit for entanglement preservation at 52 m, whereas the teleportation fidelity reaches the classical limit at 49 m. 

As we can observe, microwave quantum communication is highly limited by diffraction and thermal noise. However, inside the atmosphere, the attenuation suffered by microwaves in severe weather conditions is inferior to that suffered by optical signals. Let us look at an example, and compare the performance of signals in both regimes. To account for the effects of rain on atmospheric attenuation and visibility, we set $\alpha_{0} = 3.4\times 10^{-4} \text{ m}^{-1}$~\cite{Kaushal2017} and, in the Hufnagel-Valley turbulence model, we now write $A_{\text{day(night)}}=3.15 (2.15)\times10^{-14} \text{ m}^{-2/3}$~\cite{Liorni2019}. This exemplifies adverse meteorological conditions for optical signals, which is a convenient scenario for a comparison between microwave and optical. In the microwave regime, we need to set the water-vapor density to $p_{0} = 12 \text{ g/m}^{3}$~\cite{Ho2004}.

We observe that, when the link starts on the ground, microwaves can only do as well as optical for a short distance. This can be observed in Fig.~\ref{fig6}, where we represent the negativity (a) and the teleportation fidelity (b) associated to a TMSV state distributed through free space. The effects of diffraction remain severe on microwave signals. These results show that microwave quantum communication can be appropriate for inter-satellite quantum communications. There, the conditions for entanglement preservation become $\tau>0.706$ for asymmetric states, and $\tau>0.847$ for symmetric ones, with an effective number of thermal photons $n = 2.39$. 

\begin{figure*}[t]
{\includegraphics[width=0.95 \textwidth]{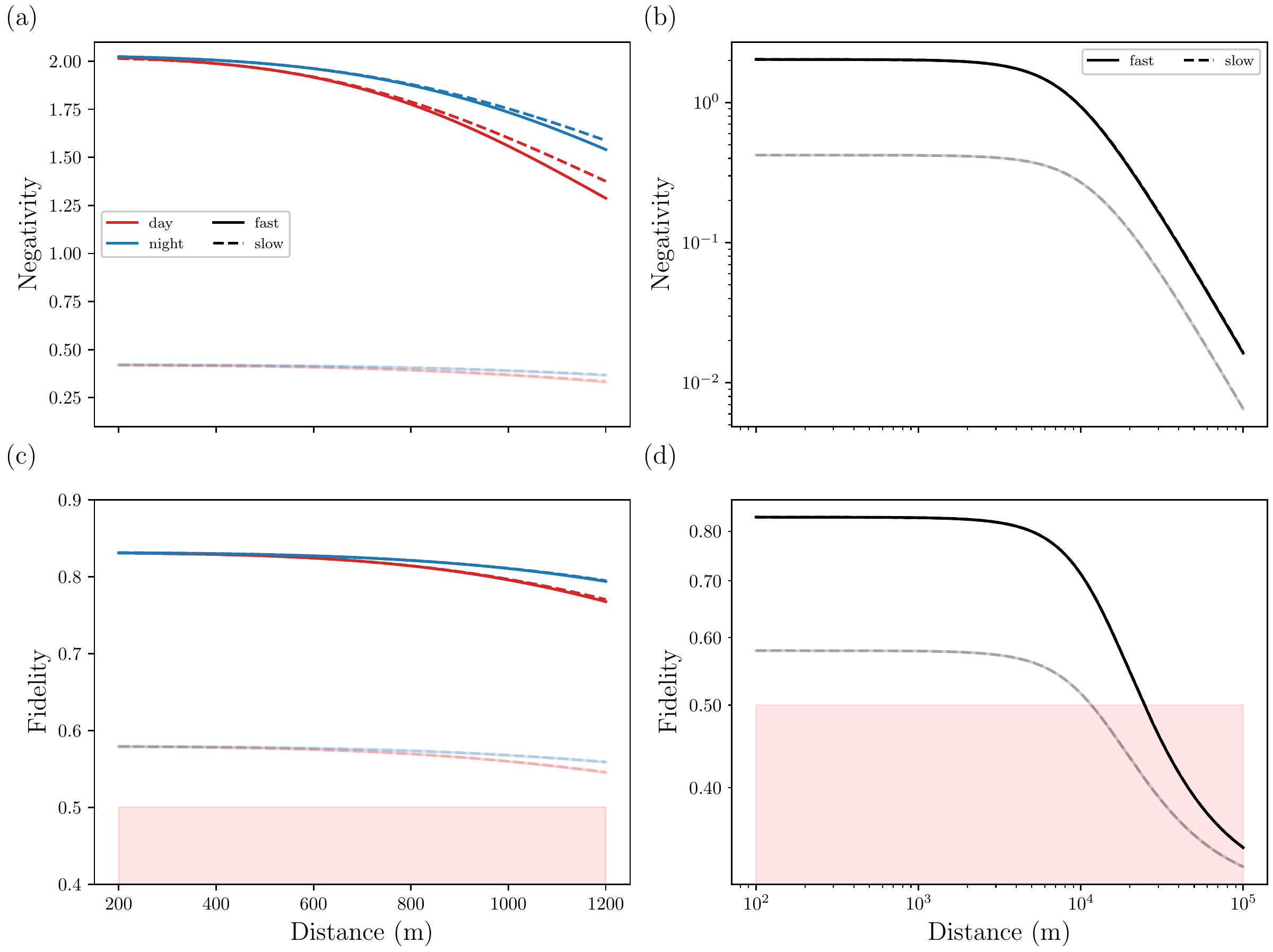}}
\caption{Quantum communication through horizontal paths with TMSV states distributed through free space. Ground-to-ground station quantum communication is studied through the negativity~(a) and the quantum teleportation fidelity~(c), for TMSV states subject to a loss mechanism composed of diffraction, atmospheric extinction, detector inefficiency and free-space turbulence. Inter-satellite quantum communication is studied through the negativity~(b) and the quantum teleportation fidelity~(d), where now the only loss mechanisms relevant are diffraction, pointing errors, and detector inefficiency. We consider the signal has wavelength $\lambda=800$ nm, squeezing parameter $r=1$, and initial waist $\varpi_{0}=5$ cm, assuming the receiver has an antenna of radius $a_{R}=5$ cm. In red, we represent the results that incorporate daytime thermal noise, whereas the blue lines consider nighttime thermal noise. The dashed lines correspond to the instance of slow turbulence and fast detection, and the solid lines correspond to fast turbulence and slow detection. In full color, we present the results for perfect detector efficiency, $\tau_{\text{eff}}=1$, whereas the transparent curves correspond to $\tau_{\text{eff}}=0.4$.}
\label{fig7}
\end{figure*}

\section{Horizontal paths}
For the sake of completeness, we investigate the effects that free-space propagation through turbulent media inside the atmopshere has on the negativity of TMSV states, and how it affects the fidelity of a quantum teleportation protocol that uses these states as resources, in order to teleport an unknown coherent state. We consider a scenario in which TMSV states are distributed between two ground stations, at an altitude of $h=30$ m, each station having an receiving antenna with $a_{R}=5$ cm of aperture radius, and able to generate quasi-monochromatic beams with wavelength $\lambda= 800$ nm and $\varpi_{0}=5$ cm of initial waist. 

In this situation, since the altitude is fixed, and for a wind speed of $v=21 \text{ m}/\text{s}$, the refraction index structure constant is $C_{n}^{2}=2.06(1.29)\times10^{-14} \text{ m}^{-2/3}$ for daytime (nighttime) values. We characterize the excess noise in the detectors by $n_{\text{day}}=4.75\times10^{-3}$ thermal photons for daytime events, and $n_{\text{night}}=4.75\times10^{-8}$ thermal photons for nighttime events.

We represent the results of entanglement distribution and quantum teleportation with TMSV states between two ground stations in Fig.~\ref{fig7}. Daytime (nighttime) results are shown in red (blue), and the solid (dashed) curves correspond to fast (slow) turbulence. The high-transparency curves show the results for inefficient detectors, with $\tau_{\text{eff}}=0.4$, whereas the curves in full color correspond to ideal detection, with $\tau_{\text{eff}}=1$. In Fig.~\ref{fig7}~(a), we show the negativity of the TMSV state, with squeezing parameter $r=1$, against the traveled distance. We show the average fidelity of quantum teleportation using these states, distributed through free space, in Fig.~\ref{fig7}~(b). We observe that, even in the low detector-efficiency case, entanglement is preserved, and therefore quantum teleportation fidelity is still higher than the maximum classical fidelity achievable, marked in a pale red background in Fig.~\ref{fig7}~(b).

The range of distances chosen to represent these quantities corresponds to the ``sweet spot'' $200\leq z\leq1066$, were the weak-turbulence expansion used here is approximately correct~\cite{Pirandola2021_2}. 

In Fig.~\ref{fig7}~(b) and~(d), we represent the negativity~(b) and the quantum teleportation fidelity~(d) for the same TMSV states, between two satellites in the same orbit. In this scenario, the only relevant sources of noise are diffraction, pointing errors, and detector inefficiency. Also, we are considering that the excess noise in the detectors is characterized by $n=8.48\times10^{-9}$ thermal photons. Here, the solid lines are associated to fast turbulence and slow detection, whereas the dashed lines describe slow-turbulence and fast-detection results. Notice that these appear overlapped. A similar study regarding microwave signals can be found in Ref.~\cite{GonzalezRaya2022}. There, the size of the antenna is taken to be much larger. Given the long wavelengths, the size of the antenna, as well as the initial spot size of the beam are crucial parameters for entanglement preservation.

\section{Conclusions}
We have studied the effects of diffraction, atmospheric attenuation, detector inefficiency and turbulence on quantum signals propagating through free space, between a ground station and a satellite. More precisely, we have investigated the effects of these loss mechanisms combined, and described as a fading quantum channel, acting on TMSV states, which are a paradigmatic example of entangled Gaussian quantum states. We have observed the degradation of entanglement through the negativity of the state, and looked at the fidelity of performing quantum teleportation with the remaining entangled resource, both after downlink and uplink communications, and for satellites in different orbits. We conclude that the best case occurs when we use a downlink, i.e. when the bipartite states are generated in the satellite and one of the modes is sent down to the ground station. Downlink quantum teleportation can be performed in 400 km before reaching maximum classical fidelities. The uplink represents the worst case because turbulence effects, which are more drastic inside the atmosphere, distort the waist of the beam and displace the focusing point; when considering the whole path, these errors have a higher impact on a beam that is starting its path. 

We have also considered the introduction of an intermediate station; we first investigated a scenario in which the states were generated there, and one mode was then sent to the ground station through a downlink, while the other was sent to the satellite through an uplink. Considering that now the uplink does not start inside the atmosphere, the results for the negativity were slightly better than those for the downlink in the simple case, provided an optimal placement of the intermediate station. Furthermore, the results for the fidelity were highly improved because the generation of states in an intermediate station leads to states that are almost symmetric. As we discussed in Appendix~\ref{app_C}, for two Gaussian states with the same entanglement, the one that presents a more symmetric covariance matrix will have a higher teleportation fidelity, in the well-known Braunstein-Kimble quantum teleportation protocol. This fidelity reaches the maximum classical value for links going into the LEO region. The second intermediate-station scenario we considered was one were the beam could be refocused, but in a simple downlink or uplink. The uplink showed a higher improvement than the downlink, because a refocusing station can help mitigate the combined effects of diffraction and turbulence, which as we discussed earlier, are more severe on more ideal beams.  

We have followed by studying a similar free-space loss mechanism for microwave signals, which are largely affected by diffraction and thermal noise. Although atmospheric absorption and turbulence effects can be neglected, the distances for entanglement distribution and effective quantum teleportation were highly reduced with respect to the optical case. In a bad weather scenario, we observed that microwave and optical signals yield a similar performance for short distances, microwaves then leading to worse results as we separate from the source, mainly due to diffraction.

We have concluded by showing the limits of entanglement distribution and quantum teleportation through horizontal paths, in ground-to-ground scenarios, where turbulence effects are present, and inter-satellite quantum communication, where we have mostly diffraction and pointing errors. Between satellites, the loss mechanism is reduced to diffraction and beam wandering, and therefore entanglement and quantum teleportation fidelity can be preserved for longer distances than in horizontal paths between ground stations, where atmospheric absorption and turbulence come into play.

We believe these results are relevant in the development of quantum communications in free space, establishing limits based on experimental parameters to the realizability of quantum information transfer between Earth and satellites. Provided that classical secure channels are well-established, the installation of adequate quantum channels represents the next step towards developing quantum communication networks, that can lead to advances such as the quantum internet.

\acknowledgements
TG-R and MS acknowledge financial support from the Basque Government through Grant No. IT1470-22 and from the Basque Government QUANTEK project under the ELKARTEK program (KK-2021/00070), the Spanish Ram\'{o}n y Cajal Grant No. RYC-2020-030503-I, project Grant No. PID2021-125823NA-I00 funded by MCIN/AEI/10.13039/501100011033 and by ``ERDF A way of making Europe'' and ``ERDF Invest in your Future'', as well as from the project QMiCS (Grant No. 820505) and the HORIZON-CL4-2022-QUANTUM-01-SGA project 101113946 OpenSuperQPlus100 of the EU Flagship on Quantum Technologies, and the EU FET-Open projects Quromorphic (828826) and EPIQUS (899368).

S.P. Acknowledges funding from the EU via CiViQ (grant agreement no. 820466) and QUARTET (Grant Agreement No. 862644), and EPSRC via the Quantum Communications Hub (Grant number EP/T001011/1).

\appendix

\section{Fast-turbulence fading channel}\label{app_A}
Let us look at how to derive the average transmissivity of a Gaussian attenuation channel to describe a fast fading channel. We will do this using the covariance-matrix formalism. First, see that the Wigner function of the state that results from applying the fading channel is
\begin{equation}
W'(x,p) = \int_{0}^{\tau_{\text{max}}}d\tau P(\tau)W_{\tau}(x,p).
\end{equation}
Here, the Wigner function $W_{\tau}(x,p)$ results from the modification of the quadrature operators $\hat{x}$ and $\hat{p}$ by the quantum channel instance $\varepsilon_{\tau}$,
\begin{eqnarray}
\nonumber \hat{x} &\rightarrow& \hat{x}_{\tau} = \sqrt{\tau}\hat{x} + \sqrt{1-\tau}\hat{x}_{\text{e}}, \\
\hat{p} &\rightarrow& \hat{p}_{\tau} = \sqrt{\tau}\hat{p} + \sqrt{1-\tau}\hat{p}_{\text{e}},
\end{eqnarray}
which get mixed with the quadrature operators of the state of an environment. In this formalism, the expectation value of the operator $\hat{A}\hat{B}$ is computed as
\begin{eqnarray}
\nonumber && \langle\hat{A}\hat{B}\rangle = \int \text{d}x\text{d}p A B W'(x,p) \\
\nonumber &&=  \int_{0}^{\tau_{\text{max}}}d\tau P(\tau) \int \text{d}x\text{d}p A B W_{\tau}(x,p) \\
&&= \int_{0}^{\tau_{\text{max}}}d\tau P(\tau)\langle\hat{A}\hat{B}\rangle_{\tau}.
\end{eqnarray}
This result implies that we can replace the elements of the covariance matrix of the state resulting from the fading channel by the weighted integral of the expectation values resulting from each channel instance~\cite{Dong2010,Usenko2012}. The later looks as follows, for the second moments of quadrature operators:
\begin{eqnarray}
\nonumber \hat{x}^{2} &\rightarrow& \hat{x}_{\tau}^{2} = \tau\hat{x}^{2} + (1-\tau)\hat{x}_{\text{e}}^{2} + \sqrt{\tau(1-\tau)}\{\hat{x},\hat{x}_{\text{e}}\}, \\
\nonumber \hat{p}^{2} &\rightarrow& \hat{p}_{\tau}^{2} = \tau\hat{p}^{2} + (1-\tau)\hat{p}_{\text{e}}^{2} + \sqrt{\tau(1-\tau)}\{\hat{p},\hat{p}_{\text{e}}\}, \\
\nonumber \{\hat{x},\hat{p}\} &\rightarrow& \{\hat{x}_{\tau},\hat{p}_{\tau}\} = \tau\{\hat{x},\hat{p}\} + (1-\tau)\{\hat{x}_{\text{e}},\hat{p}_{\text{e}}\} \\
&+& \sqrt{\tau(1-\tau)}\left(\{\hat{x},\hat{p}_{\text{e}}\} + \{\hat{p},\hat{x}_{\text{e}}\}\right).
\end{eqnarray}
For the complete fading channel, we will have to make the replacement
\begin{eqnarray}
\nonumber \tau &\longrightarrow& \langle\tau\rangle = \int_{0}^{\tau_{\text{max}}}d\tau P(\tau)\tau, \\
\sqrt{\tau} &\longrightarrow& \left\langle\sqrt{\tau}\right\rangle = \int_{0}^{\tau_{\text{max}}}d\tau P(\tau)\sqrt{\tau}.
\end{eqnarray}

\section{Intermediate station improvement}\label{app_C}
\begin{figure*}[t]
{\includegraphics[width=0.95 \textwidth]{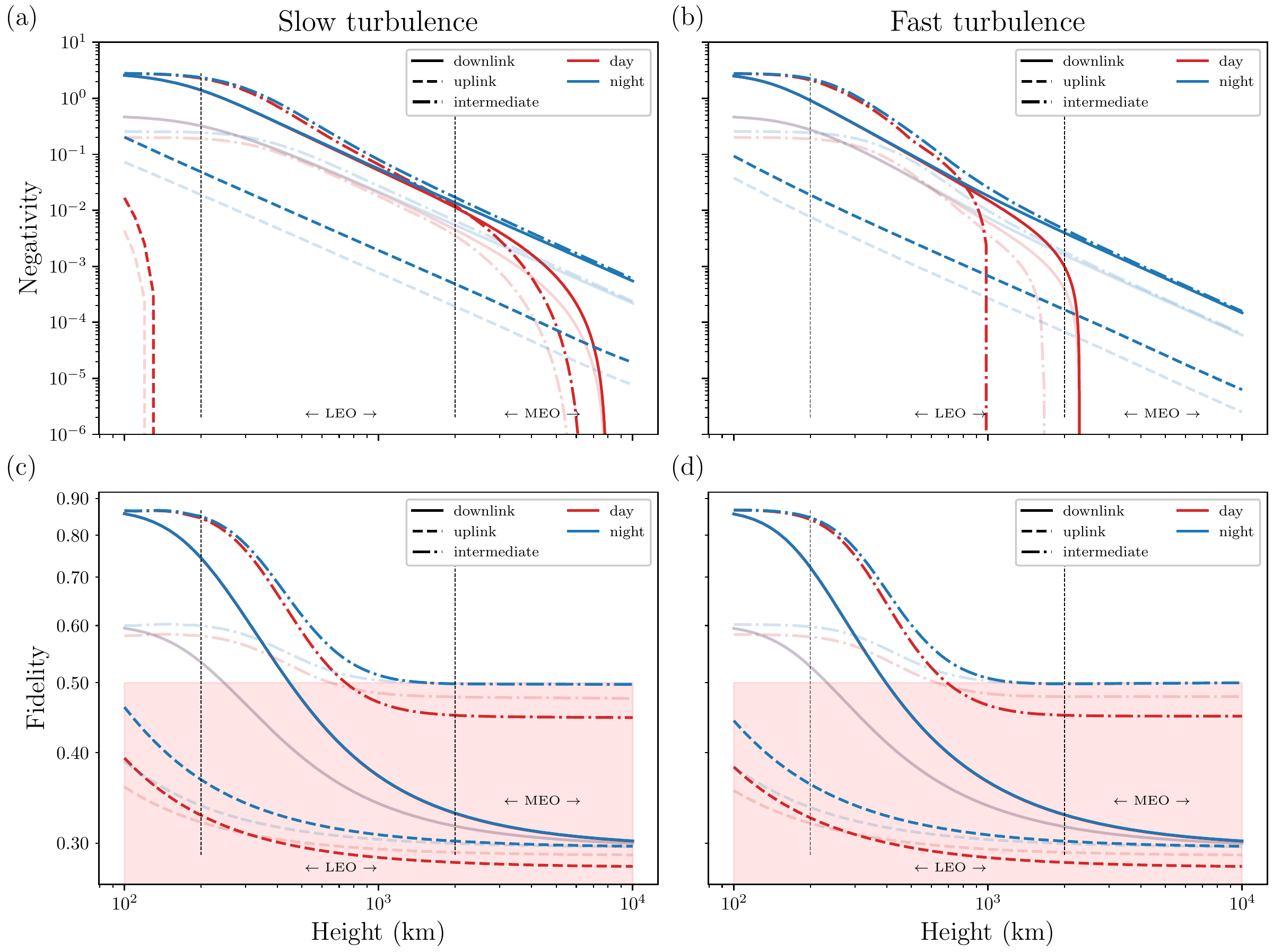}}
\caption{Comparison between the use of a downlink or an uplink to distribute two-mode entangled Gaussian quantum states between a ground station and a satellite, and the use of an intermediate station, where these states are generated. We study a signal of wavelength $\lambda=800$ nm and initial waist $\varpi_{0}=20$ cm, in a TMSV state with squeezing parameter $r=1$, sent to a receiver that has an antenna of radius $a_{R}=40$ cm, and that is subject to a loss mechanism composed of diffraction, atmospheric extinction, detector inefficiency and free-space turbulence, and described by a fading channel. In a solid line, we plot the quantities associated to a downlink; those corresponding to an uplink appear dashed, and the dashed-dotted lines describe the case of an intermediate station. In full color we represent the results for perfect detector efficiency, $\tau_{\text{eff}}=1$, whereas imperfect detection, $\tau_{\text{eff}}=0.4$, is marked by the transparent curves. Negativity of the state after the fading channel in the slow-turbulence and fast-detection regime (a), and in the fast-turbulence and slow-detection regime (b), against the height of the complete link. The average fidelity of teleporting an unknown coherent state, using the entangled states that result from the fading process, in the slow-turbulence and fast-detection regime (c), and in the fast-turbulence and slow-detection regime (d), against the height of the complete link.}
\label{fig8}
\end{figure*}

Here, we elaborate the discussion regarding the improvements that the generation of bipartite states at an intermediate station can bring, for Gaussian states propagating through turbulent media. In Fig.~\ref{fig8}, we present the different negativities and fidelities, for fast and slow turbulence regimes. We compare the case of a downlink, an uplink, and the combination required by an intermediate station, against the height of the link. We can observe in Fig.~\ref{fig8}~(a), (b) that the negativity, in the case of an intermediate station is larger that a single dowlink/uplink, but only in the ideal case; when we consider inefficient detectors ($\tau_{\text{eff}}=0.4$), this gain is not so clear. As we increase the height of the link, this gain is not significant with respect to the downlink, although it remains relevant against the uplink. In Fig.~\ref{fig8}~(c), (d) the fidelity of teleporting an unknown coherent state is much better with an intermediate station, with respect to either a downlink or an uplink. Specially, we can highlight its partial saturation at the maximum classical fidelity value. 

\begin{figure}[t]
{\includegraphics[width=0.48 \textwidth]{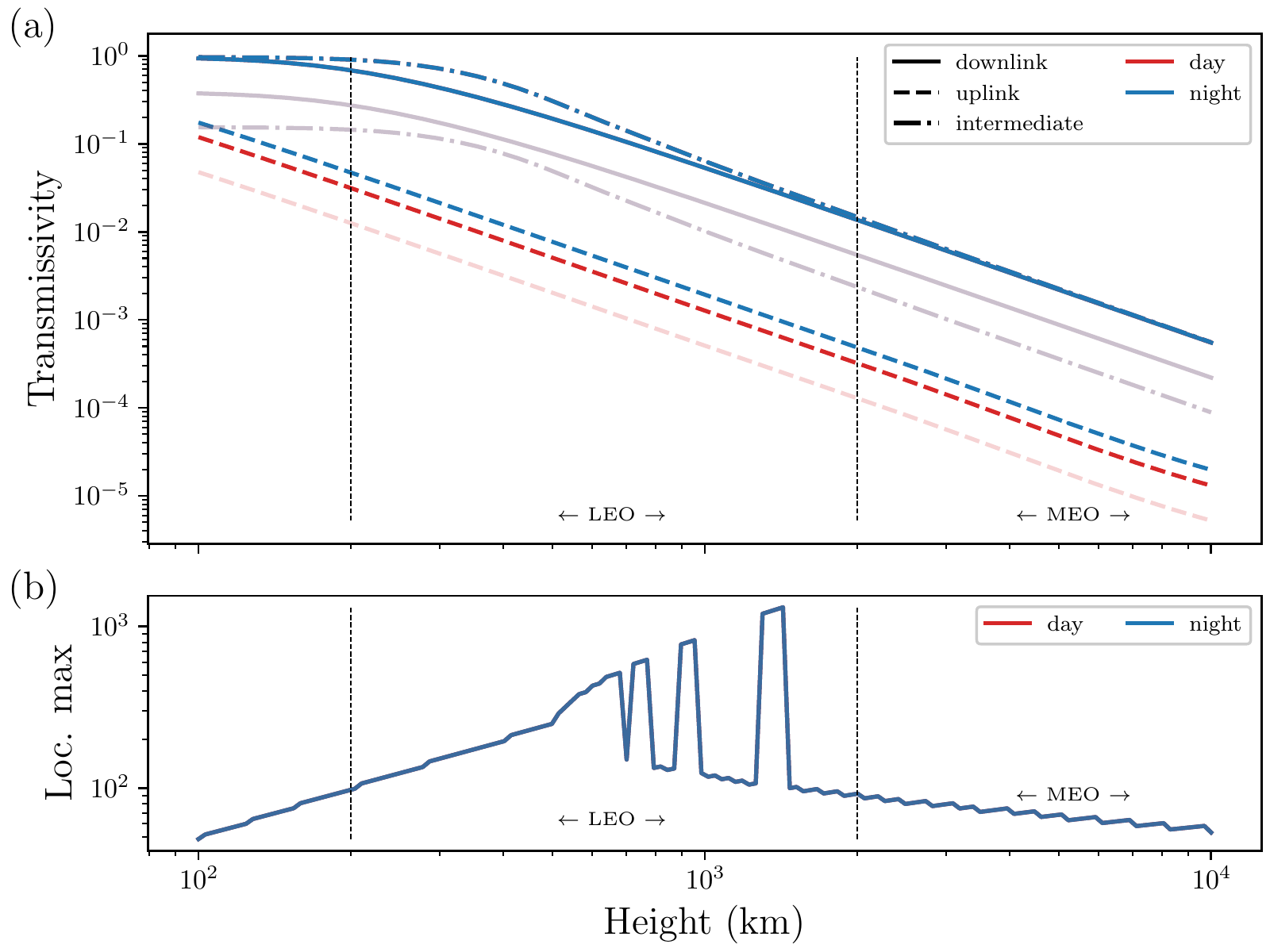}}
\caption{(a) Average transmissivity of a fading channel modelling a loss mechanism present in a link connecting a ground station and a satellite, and composed of diffraction, atmospheric extinction, detector inefficiency and free-space turbulence. In a solid line, we plot the quantities associated to a downlink; those corresponding to an uplink appear dashed, and the dashed-dotted lines describe the case of an intermediate station. In full color we represent the results for perfect detector efficiency, $\tau_{\text{eff}}=1$, whereas imperfect detection, $\tau_{\text{eff}}=0.4$, is marked by the high-transparency curves. In blue, we represent the result associated with nighttime thermal noise, whereas those associated to daytime thermal noise appear in red. (b) Optimal positions of the intermediate station, in order to maximize the transmissivity, against the height of the link.}
\label{fig9}
\end{figure}

We observe that the transmissivity in the case of the intermediate station is only higher than that of the downlink in the ideal case; when we have imperfect detectors, since there are now two detection events, the transmissivity is always worse. This can be observed in Fig.~\ref{fig9}, where we represent the transmissivity induced by downlink, uplink, and intermediate-station scenarios. We consider nighttime and daytime noise, again seeing that transmissivities associated to downlink and intermediate-station communication coincide. The same thing happens in Fig.~\ref{fig8}. Although we can see this behavior in the negativity plots, the fidelity behaves different. This is because the state is more symmetric in the case of the intermediate station, and this improves the teleportation fidelity, such that for a perfectly symmetric state, the teleportation fidelity if $\overline{F}=1/(1+\tilde{\nu}_{-})$, given a partially-transposed symplectic eigenvalue $\tilde{\nu}_{-}$ that completely characterizes the negativity. 

Take a covariance matrix like the one in Eq.~\eqref{CM}, assuming it represents an asymmetric state. Here, we are referring to symmetry in the second moments of modes A and B, and not in the sense that the covariance matrix is symmetric. The partially-transposed symplectic eigenvalue is 
\begin{equation}
\tilde{\nu}_{-}^{\text{A}} = \frac{\alpha+\beta-\sqrt{(\alpha-\beta)^{2}+4\gamma^{2}}}{2}
\end{equation}
and the teleportation fidelity is
\begin{equation}
\overline{F}_{\text{A}} = \frac{1}{1+(\alpha+\beta-2\gamma)/2}.
\end{equation}
For a symmetric Gaussian state with covariance matrix 
\begin{equation}
\Sigma_{\text{S}} = \begin{pmatrix} \delta\mathbb{1}_{2} & \varepsilon Z \\ \varepsilon Z & \delta\mathbb{1}_{2} \end{pmatrix},
\end{equation}
the partially-transpose symplectic eigenvalue is
\begin{equation}
\tilde{\nu}_{-}^{\text{S}} = \delta-\varepsilon
\end{equation}
and the teleportation fidelity is
\begin{equation}
\overline{F}_{\text{S}} = \frac{1}{1+\delta-\varepsilon}.
\end{equation}
If these two states have the same negativity, then
\begin{equation}
\frac{\alpha+\beta-\sqrt{(\alpha-\beta)^{2}+4\gamma^{2}}}{2} = \delta - \varepsilon,
\end{equation}
and we can write 
\begin{equation}
\overline{F}_{\text{S}} = \frac{1}{1+\frac{\alpha+\beta-\sqrt{(\alpha-\beta)^{2}+4\gamma^{2}}}{2}}.
\end{equation}
Claiming that the fidelity with the symmetric state is higher than that with the asymmetric state amounts to checking that
\begin{equation}
\alpha+\beta-\sqrt{(\alpha-\beta)^{2}+4\gamma^{2}} < \alpha+\beta-2\gamma,
\end{equation}
and this is always true for $\alpha\neq\beta$. This statement works for a perfectly symmetric state, but we can write an extension for more general covariance matrices. We take
\begin{equation}
\Sigma_{1} = \begin{pmatrix} \alpha_{1}\mathbb{1}_{2} & \gamma_{1} Z \\ \gamma_{1} Z & \beta_{1}\mathbb{1}_{2} \end{pmatrix}, \qquad \Sigma_{2} = \begin{pmatrix} \alpha_{2}\mathbb{1}_{2} & \gamma_{2} Z \\ \gamma_{2} Z & \beta_{2}\mathbb{1}_{2} \end{pmatrix},
\end{equation}
and assume that $\alpha_{i}>\beta_{i}$ ($i\in\{1,2\}$) for convenience, and expand up to first order in $\alpha_{i}-\beta_{i}\ll 1$. We can say that, if the states represented by these two covariance matrices have the same negativity, then state 1 shows higher teleportation fidelity for an unknown coherent state if
\begin{equation}
\alpha_{1}-\beta_{1} < \sqrt{\frac{\gamma_{1}}{\gamma_{2}}}(\alpha_{2}-\beta_{2}).
\end{equation}
This also works the other way around; for two states with the same teleportation fidelity, state 1 shows lower entanglement if its covariance matrix elements satisfy the above condition. Furthermore, we could fix $\gamma_{1}=\gamma_{2}=\gamma$, and see that for higher orders of the expansion $\alpha_{i}-\beta_{i}\ll 1$, we obtain that $\alpha_{1}-\beta_{1} < \alpha_{2}-\beta_{2}$ if 
\begin{equation}
\sqrt{(\alpha_{1}-\beta_{1})^{2} +  (\alpha_{2}-\beta_{2})^{2}} < 4\gamma 
\end{equation}
is satisfied. 

\end{document}